\documentclass[pra,twocolumn,preprintnumbers,amsmath,amssymb,nofootinbib,floatfix,superscriptaddress]{revtex4}

\usepackage{graphicx, bm, tikz}
\usepackage[breaklinks]{hyperref}

\makeatletter
\def\graphicscale{\twocolumn@sw{0.3}{0.4}}
\def\graphicthreescale{\twocolumn@sw{0.3}{0.4}}

\begin{document}

\title{Dissipative dynamics at first-order quantum transitions}

\author{Giovanni Di Meglio}
\affiliation{Dipartimento di Fisica dell'Universit\`a di Pisa, 
Largo Pontecorvo 3, I-56127 Pisa, Italy}

\author{Davide Rossini}
\affiliation{Dipartimento di Fisica dell'Universit\`a di Pisa
        and INFN, Largo Pontecorvo 3, I-56127 Pisa, Italy}

\author{Ettore Vicari} 
\affiliation{Dipartimento di Fisica dell'Universit\`a di Pisa
        and INFN, Largo Pontecorvo 3, I-56127 Pisa, Italy}

\date{\today}

\begin{abstract}
  We investigate the effects of dissipation on the quantum dynamics of
  many-body systems at quantum transitions, especially considering
  those of the first order.  This issue is studied within the
  paradigmatic one-dimensional quantum Ising model.  We analyze the
  out-of-equilibrium dynamics arising from quenches of the Hamiltonian
  parameters and dissipative mechanisms modeled by a Lindblad master
  equation, with either local or global spin operators acting as
  dissipative operators.  Analogously to what happens at continuous
  quantum transitions, we observe a regime where the system develops a
  nontrivial dynamic scaling behavior, which is realized when the
  dissipation parameter $u$ (globally controlling the decay rate of
  the dissipation within the Lindblad framework) scales as the energy
  difference $\Delta$ of the lowest levels of the Hamiltonian, i.e.,
  $u\sim \Delta$.  However, unlike continuous quantum transitions
  where $\Delta$ is power-law suppressed, at first-order quantum
  transitions $\Delta$ is exponentially suppressed with increasing the
  system size (provided the boundary conditions do not favor any
  particular phase).
\end{abstract}

\maketitle


\section{Introduction}
\label{sec:Introduction}

The recent progress in atomic physics and quantum optical technologies
has enabled great opportunities for a thorough investigation of the
interplay between the coherent quantum dynamics and the (practically
unavoidable) dissipative effects, due to the interaction with an
external environment~\cite{HTK-12, MDPZ-12, CC-13, AKM-14}.  The
competition between coherent and dissipative dynamic mechanisms may
originate a nontrivial interplay, which can be responsible for the
emergence of further interesting phenomena in many-body systems, in
particular close to a quantum phase transition where the many-body
systems develop peculiar quantum correlations~\cite{Sachdev-book}.

Certain issues related to the competition between coherent and
dissipative dynamics have been addressed at continuous quantum
transitions (CQTs)~\cite{YMZ-14, YLC-16, NRV-19, RV-19, RV-20}, where
quantum correlations develop diverging length scales $\xi$, and the gap
$\Delta$ closes as a power law of $\xi$, i.e. $\Delta\sim\xi^{-z}$,
$z$ being the universal dynamic exponent.  These studies considered a
class of dissipative mechanisms which can be reliably described by a
Lindblad master equation governing the time evolution of the system's
density matrix.  It was argued, and numerically checked, that a
dynamic scaling limit exists at a CQT even in the presence of
dissipation, whose main features are controlled by the universality
class of the quantum transition.  However, such a dynamic scaling
limit requires a particular tuning of the dissipative interactions,
whose decay rate $u$ should scale as $u\sim \Delta \sim \xi^{-z}$.

In this paper we extend the above studies to first-order quantum
transitions (FOQTs), which have their own peculiarities, in particular
related to the emergence of an exponentially suppressed gap and to
their sensitivity to the boundary conditions in finite
systems~\cite{CNPV-14, PRV-18-fo, PRV-20}.  Besides that, FOQTs are of
great phenomenological interest, since they occur in a large variety
of many-body systems, including quantum Hall samples~\cite{PPBWJ-99},
itinerant ferromagnets~\cite{VBKN-99}, heavy-fermion
metals~\cite{UPH-04, Pfleiderer-05, KRLF-09}, disordered
systems~\cite{JKKM-08, JLSZ-10}, and infinite-range
models~\cite{YKS-10, SN-12}.

\begin{figure}[!t]
  \includegraphics[width=0.9\columnwidth]{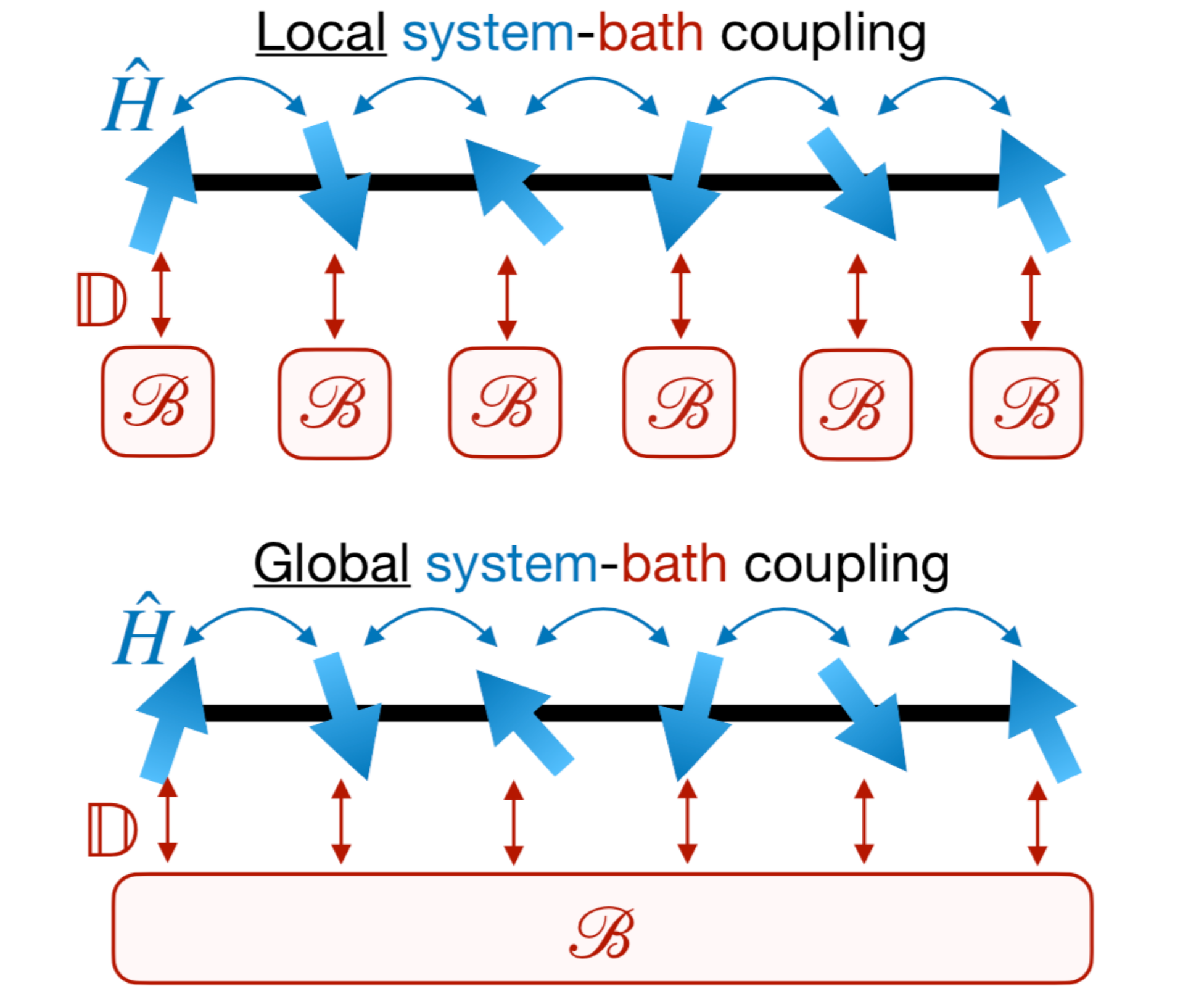}
  \caption{The quantum spin-chain model discussed in this work.
    Neighboring spins are coupled through a coherent Hamiltonian $\hat
    H$ (bidirectional blue arrows). Each spin is also homogeneously
    and weakly coupled to some external bath ${\cal B}$ via a set of
    dissipators $\mathbb{D}$ (vertical red arrows), whose effect is to
    induce incoherent dissipation.  The environment is modeled either
    as a sequence of local independent baths, each for any spin of the
    chain (top drawing), or as a single common bath to which each spin
    is supposed to be uniformly coupled (bottom drawing).}
  \label{fig:sketch}
\end{figure}

We address the interplay between the critical coherent dynamics and
dissipative mechanisms, when the Hamiltonian parameters are close to a
FOQT.  To this purpose, we consider dynamic protocols that start from
ground states close to FOQTs, and then analyze the out-equilibrium
dynamics arising from a instantaneous quench of the Hamiltonian
parameters and the dissipative interaction with the environment.  We
take, as a paradigmatic example, the one-dimensional spin-$1/2$
quantum Ising model, exhibiting a FOQT line in its zero-temperature
phase diagram, in the presence of either local or global homogeneous
dissipative mechanisms---see Fig.~\ref{fig:sketch}.  Their effects are
assumed to be well captured by a Lindblad master equation of the
density matrix of the open system.  We mention that
the dissipative dynamics of spin models at first-order transitions
has been recently considered in Ref.~\cite{NF-20}, discussing
various variants of the Lindblad equations.

Analogously to what happens at CQTs~\cite{NRV-19, RV-19}, the quantum
Ising chain along the FOQT line unveils a regime where a nontrivial
dynamic scaling behavior is developed. This is observed when the
dissipation parameter $u$ (globally controlling the decay rate of the
dissipation within the master Lindblad equation) scales as the energy
difference $\Delta$ of the lowest levels of the Hamiltonian of the
many-body system, i.e., $u\sim \Delta$.  However, unlike CQTs where
$\Delta$ is power-law suppressed, at FOQTs $\Delta$ is exponentially
suppressed with increasing the system size (when the boundary
conditions do not favor any particular phase).  The dynamic scaling
behavior turns out to become apparent for relatively small systems
already, such as chains with $L\lesssim 10$.  This makes such dynamic
scaling phenomena particularly interesting even from an experimental
point of view, where the technical difficulties in manipulating and
controlling such systems can be probably faced with up-to-date
methods.

Here the out-of-equilibrium quantum dynamics associated with the
above-mentioned protocol is numerically monitored by considering
standard observables, such as the longitudinal magnetization, as well
as the average work and heat characterizing the quantum thermodynamic
properties of the out-of-equilibrium phenomenon.

The paper is organized as follows.  In Sec.~\ref{model} we introduce
the one-dimensional quantum Ising chain, the modelization of the
dissipative interactions through a master Lindblad equation, and the
out-of-equilibrium protocol discussed hereafter.
Sec.~\ref{dynbehsumm} presents a summary of the main features of the
dynamic scaling theory for closed systems, and their extension to
allow for dissipative interactions; in particular we address the
behavior expected at FOQTs.  In Sec.~\ref{foqtres} we numerically
study the dynamics of quantum Ising chains arising from the
above-mentioned protocol, up to size of order $L=10$, along the FOQTs
and at the CQT, showing that the results support the general dynamic
scaling theory.  Finally in Sec.~\ref{conclu} we summarize and draw
our conclusions.  In App.~\ref{1spinModel} we solve the analogous
problem for a single quantum spin. In App.~\ref{kitaev} we address
some related questions for the out-of-equilibrium behavior of the
Kitaev fermionic wire (related to the quantum Ising chain by a
Jordan-Wigner mapping) in the presence of local dissipation due to
particle pumping or decay, at its CQT; these include exact analytic
results for the quantum work and the heat interchange during its time
evolution.

\section{The out-of-equilibrium protocol}
\label{model}

The Hamiltonian of the one-dimensional ferromagnetic quantum Ising
model, reads:
\begin{equation}
  \hat H(g,h) = - J \, \sum_{x=1}^L \left[ \hat \sigma^{(1)}_{x\phantom{1}}
    \hat \sigma^{(1)}_{x+1} + g \, \hat \sigma^{(3)}_x + h \, \hat
    \sigma^{(1)}_x \right] \,,
  \label{hedef}
\end{equation}
where $\hat\sigma_x^{(k)}$ are spin-$1/2$ Pauli matrices associated
with the $x$th site of the chain, $J>0$ sets the energy scale, and we
assume a transverse field with strength $g\geq 0$. We also consider
spin systems of size $L$ with periodic boundary conditions.

It is well known that, at $g=g_c=1$ and $h=0$, the model undergoes a
CQT separating a disordered phase ($g>1$) from an ordered ($g<1$)
one~\cite{Sachdev-book}. The corresponding quantum critical behavior
belongs to the two-dimensional Ising universality class, characterized
by a diverging length scale ($\xi\sim |g-g_c|^{-\nu}$, with $\nu=1$)
and the power-law suppression of the gap $\Delta$ between the two
lowest energy levels (as $\Delta \sim \xi^{-z}$ with $z=1$, or as
$\Delta \sim L^{-z}$ in finite-size systems at the critical point).

The FOQT line, located at $g<g_c=1$, is related to the level crossing
of the two lowest states $| \! \uparrow \rangle$ and $| \! \downarrow
\rangle$ for $h=0$, such that $\langle \uparrow \! | \hat
\sigma_x^{(1)} | \! \uparrow \rangle = m_0$ and $\langle \downarrow \!
| \hat \sigma_x^{(1)} | \! \downarrow \rangle = -m_0$ (independently
of $x$), with $m_0>0$.  The degeneracy of these states is lifted by
the longitudinal field, with strength $h$.  Therefore, $h = 0$ is a
FOQT point, where the ground-state longitudinal magnetization
\begin{equation}
  M = \big\langle \hat\Sigma^{(1)} \big\rangle \,,\qquad 
  \hat \Sigma^{(1)} \equiv {1\over L} \sum_x \hat\sigma_x^{(1)} \,,
  \label{lmagnetization}
\end{equation}
becomes discontinuous in the infinite-volume limit.  The transition
separates two different phases characterized by opposite values of the
magnetization $m_0$, i.e.~\cite{Pfeuty-70}
\begin{equation}
  \lim_{h \to 0^\pm} \lim_{L\to\infty} M = \pm m_0\,, \qquad
  m_0 = \big( 1 - g^2 \big)^{1/8}\,.
  \label{m0def}
\end{equation}
In a finite system of size $L$ with either periodic or open boundary
conditions (which do not favor any of the two magnetized phases
separated by the FOQT), the lowest states are superpositions of the
two magnetized states $| \! \uparrow \rangle$ and $| \! \downarrow
\rangle$. Due to tunneling effects, the energy gap at $h=0$ vanishes
exponentially as $L$ increases~\cite{PF-83, CNPV-14}, as $\Delta_L
\sim e^{-c L^d}$.  In particular, for the one-dimensional case of
the model~\eqref{hedef}, the gap at $h=0$ behaves as~\cite{Pfeuty-70, CJ-87}
$\Delta_L \approx 2 \, (1-g^2) g^L$ for open boundary conditions, and
\begin{equation}
  \Delta_L \approx 2 \, \sqrt{1-g^2\over \pi L} \, g^L
  \label{deltapbc}
\end{equation}
for periodic boundary conditions.  The differences $E_i-E_0$ for the
higher excited states $(i>1$) are finite for $L\to \infty$.

We model the dissipative interaction with the environment by Lindblad
master equations for the density matrix of the system~\cite{BP-book,
  RH-book},
\begin{equation}
  {\partial\rho\over \partial t} = -{i\over \hslash}\, [ \hat H,\rho]
  + u \, {\mathbb D}[\rho]\,,
  \label{lindblaseq}
\end{equation}
where the first term in the r.h.s.~provides the coherent driving,
while the second term accounts for the coupling to the environment.
Its form depends on the nature of the dissipation arising from the
interaction with the bath, which is effectively described by a set of
dissipators ${\mathbb D}$, and a global coupling $u>0$.  In quantum
optical implementations, the conditions leading to such a framework to
study dissipative phenomena are typically satisfied~\cite{SBD-16},
therefore this formalism constitutes a standard choice for theoretical
investigations of such kind of systems.

In the following, we restrict to homogeneous dissipation mechanisms,
preserving translational invariance.  We mostly consider local
dissipative mechanisms such as the one sketched in the top drawing of
Fig.~\ref{fig:sketch}, whose trace-preserving superoperator ${\mathbb
  D}[\rho]$ can be written as~\cite{Lindblad-76, GKS-76}
\begin{equation}
  {\mathbb D}[\rho] 
  = \sum_{x=1}^L \hat L_x \rho \hat L_x^\dagger - \frac{1}{2}
  \big( \rho\, \hat L_x^\dagger \hat L_x + \hat L_x^\dagger \hat L_x
  \rho \big)\,.
  \label{dL}
\end{equation}
The Lindblad jump operator $\hat L_x$ associated to the local
system-bath coupling scheme is chosen to be
\begin{equation} 
  \hat L_x^{\pm} \equiv 
  \hat\sigma_x^{\pm} = \tfrac12 \big[ \hat\sigma_x^{(1)} \pm i
    \hat\sigma_x^{(2)} \big] \,,
  \label{lxdef}
\end{equation}
corresponding to mechanisms of incoherent raising ($+$) or lowering
($-$) for each spin of the chain.  We shall also consider an
alternative global dissipative interaction with the environment
(bottom drawing of Fig.~\ref{fig:sketch}), described by
\begin{equation}
  {\mathbb D}[\rho] = 
  \hat L \rho \hat L^\dagger - \frac{1}{2}
  \big( \rho\, \hat L^\dagger \hat L + \hat L^\dagger \hat L \rho \big)\,,
  \label{lglo}
\end{equation}
with a single raising, or lowering, Lindblad operator
\begin{equation}
  \hat L^{\pm} \equiv \hat\Sigma^{\pm}\,,\qquad
  \hat\Sigma^{\pm} \equiv {1\over L} \sum_x  \hat\sigma_x^{\pm} \,.
  \label{lglobdef}
\end{equation}

In the following, we address the interplay between the coherent
dynamics and dissipative mechanisms, focussing in particular on the
cases $g \leq 1$ and $|h|\ll 1$, corresponding to situations close to
the transition line.  For this purpose, we consider dynamic protocols
that start from ground states of the quantum Ising Hamiltonian close
to the transition line, and then analyze the out-equilibrium dynamics
driven by the master Lindblad equation~\eqref{lindblaseq}.
More precisely, we adopt the following protocol: (i) the system
starts, at $t=0$, from the ground state of the
Hamiltonian~\eqref{hedef} with transverse field parameter $g\leq 1$
and the longitudinal parameter $h_i$; (ii) then the system evolves
according to Eq.~\eqref{lindblaseq}, where the coherent driving is
provided by the Hamiltonian for the same value $g$ and a longitudinal
parameter $h$ (which may differ from $h_i$, giving rise to a sudden
quench), while the dissipative driving is controlled by the parameter
$u$ (for $u=0$, one recovers the unitary dynamics of closed systems).

The out-of-equilibrium evolution, for $t>0$, is monitored by measuring
certain fixed-time observables, such as the longitudinal magnetization
\begin{equation}
  M(t,h_i,h,L) = {\rm Tr} \big[ \rho(t) \, \hat\Sigma^{(1)} \big] \,,
  \label{mtdef}
\end{equation}
where the spin operator $\hat \Sigma^{(1)}$ is defined in
Eq.~\eqref{lmagnetization} and $\rho(t)$ is the density matrix of the
evolving system at time $t$. Analogously, one may also consider
fixed-time spin correlations.

We are also interested in the quantum thermodynamic properties
associated with this dissipative dynamics.  The first law of
thermodynamics describing the energy flows of the global system,
including the environment, can be written as~\cite{DC-book,
  BCGAA-book, GMM-book}
\begin{equation}
  {d E_s \over dt} = w(t) + q(t)\,,\label{filaw}
\end{equation}
where $E_s$ is the average energy of the open system
\begin{equation}
  E_s = {\rm Tr}\, \big[ \rho(t) \, \hat H(t) \big]\,,
  \label{esdef}
\end{equation}
and
\begin{eqnarray}
  w(t) & \equiv & {d W\over dt} = {\rm Tr} \bigg[ \rho(t)\, 
    {d \hat H(t)\over dt} \bigg]\,,\label{awork}\\ 
  q(t) & \equiv & {d Q\over dt} = {\rm Tr} \bigg[
    {d\rho(t)\over dt}\, \hat H(t) \bigg]  
  \,,\label{aheat}
\end{eqnarray}
with $W$ and $Q$ respectively denoting the average work done on the
system and the heat interchanged with the environment.

In our quench protocol, a nonvanishing work is only done at $t=0$,
when the longitudinal-field parameter suddenly changes from $h_i$ to
$h \neq h_i$. Since, after quenching the field, the Hamiltonian is
kept fixed [thus $w(t)=0$, for $t>0$], the average work is simply
given by the static expectation value
\begin{equation}
  W = \langle 0_{h_i} | \hat H(h) - \hat H(h_i) |
  0_{h_i}\rangle = (h_i - h ) L\, \langle 0_{h_i}
  | \hat\Sigma^{(1)} | 0_{h_i}\rangle \,,
  \label{wokih}
\end{equation}
where $|0_{h_i}\rangle$ is the starting ground state associated with
the longitudinal parameter $h_i$.  Note that the average work of this
protocol is the same of that arising at sudden quenches of closed
Ising chains, whose scaling behavior at the CQT and FOQTs has been
analyzed in Ref.~\cite{NRV-19-wo}.  On the other hand, the heat
interchange with the environment is strictly related to the
dissipative mechanism, indeed one can easily derive the relation
\begin{equation}
  q(t) = u \, {\rm Tr} \big[ {\mathbb D}[\rho]\, \hat H(t) \big]\,,
  \label{qteq}
\end{equation}
by replacing the r.h.s.~of the Lindblad equation~\eqref{lindblaseq}
into the expression~\eqref{aheat}.

\section{Dynamic scaling at quantum transitions}
\label{dynbehsumm}

We now summarize the main features of the dynamic scaling framework
that we will exploit to analyze the out-of-equilibrium quantum
dynamics of closed and open many-body systems.  The scaling hypothesis
is based on the existence of a nontrivial large-size limit, keeping
appropriate scaling variables fixed.

We focus on the quantum Ising chain~\eqref{hedef} along its transition
line, thus for $g \leq 1$ and $|h|\ll 1$, corresponding to FOQTs for
$g<1$ and CQT for $g=1$. Then we discuss the scaling behaviors arising
from a longitudinal external field $h$, i.e.
\begin{equation}
  \hat H_h = - h \sum_x \hat\sigma_x^{(1)}\,.
  \label{hhpert}
\end{equation}
The corresponding scaling variable controlling the equilibrium
properties of isolated many-body systems at both CQTs and FOQTs can be
generally written as the ratio~\cite{CNPV-14, RV-19-dec}
\begin{equation}
  \kappa(h) = E_h / \Delta_L\,,
  \label{kappah}
\end{equation}
between the $L$-dependent energy variation $E_h$ associated with the
$\hat H_h$ term and the energy difference $\Delta_L \equiv E_1-E_0$ of
the lowest-energy states at the transition point $h=0$.  Nonzero
temperatures could be taken into account as well, by adding a further
scaling variable $\tau = {T/\Delta_L}$.  Dynamic behaviors, exhibiting
nontrivial time dependencies, also require a scaling variable
associated with the time variable, which is generally given by
\begin{equation}
  \theta = \Delta_L\, t\,.
  \label{thetadef}
\end{equation}

The equilibrium and dynamic scaling limits are defined as the
large-size limit, keeping the above scaling variables fixed.  Within
this framework, the differences between CQTs and FOQTs are basically
related to the functional dependence of the above scaling variables on
the size: typically, power laws arise at CQTs, while exponential laws
emerge at FOQTs.

Specializing Eq.~\eqref{kappah} to the FOQTs of the quantum Ising
chain, we obtain the scaling variable~\cite{CNPV-14}
\begin{equation}
  \kappa(h) = 2 m_0 h L / \Delta_L \,,
  \label{kappaising}
\end{equation}
since $2m_0 h L$ quantifies the energy associated with the
corresponding longitudinal-field perturbation $\hat H_h$, and
$\Delta_L\sim g^L$.  For example, in the equilibrium finite-size
scaling limit, the magnetization is expected to behave
as~\cite{CNPV-14} $M(h,L) = m_0 \, {\cal M}(\kappa)$, where ${\cal M}$
is a suitable scaling function.  We point out that the FOQT scenario
based on the avoided crossing of two levels is not realized for any
boundary condition~\cite{CNPV-14, PRV-20}: in fact, in some cases the
energy difference $\Delta_L$ of the lowest levels may even display a
power-law dependence on $L$.  However, the scaling variable
$\kappa(h)$ obtained using the corresponding $\Delta_L$ turns out to
be appropriate, as well~\cite{CNPV-14}.  In the rest of the paper we
shall restrict our study to quantum Ising models with boundary
conditions that do not favor any of the two magnetized phases, such as
periodic or open boundary conditions, which generally lead to
exponential finite-size scaling laws.  Thus also the scaling variable
$\theta$ related to the time dependence, cf. Eq.~\eqref{thetadef}, is
subject to an exponential rescaling.

The CQT at $g=1$ and $h=0$ is characterized by power laws,
irrespective of the boundary conditions (see, e.g.,
Ref.~\cite{CPV-14}).  The corresponding scaling variable turns out to
be
\begin{equation}
  \kappa(h) \propto L^{y_h} h \,,
\end{equation}
where $y_h=15/8$ is the renormalization-group dimension of the
longitudinal field $h$ [see, e.g., Ref.~\cite{RV-19-dec} for its
  derivation from the general expression given in Eq.~\eqref{kappah}].
Moreover, the energy difference between the two-lowest states behaves
as $ \Delta_L \sim L^{-z}$ where $z=1$.

For example, let us consider a quench of the longitudinal field of a
closed quantum Ising chain at $t=0$, from $h_i$ (starting from the
corresponding ground state) to $h\neq h_i$.  At both the FOQTs and the
CQT, we expect that the quantum coherent evolution of the longitudinal
magnetization~\eqref{mtdef} develops the dynamic scaling behavior
\begin{equation}
  M(t,h_i,h,L) \approx L^{-\zeta} F_m(\theta,\kappa_i,\kappa)\,,
  \label{scamag}
\end{equation}
where 
\begin{equation}
  \kappa_i\equiv \kappa(h_i)\,,\qquad \kappa\equiv \kappa(h)\,,
  \label{defkaika}
\end{equation}
the exponent $\zeta = 1/8$ at the CQT (related to the
renormalization-group dimension of the longitudinal spin variable),
while $\zeta=0$ at the FOQTs, and $F_ M$ is an appropriate scaling
function.  The approach to such asymptotic behavior is generally
characterized by power-law corrections~\cite{CPV-14, CNPV-14,
  RV-19-dec}.

As discussed in Refs.~\cite{NRV-19, RV-19}, at CQTs the dissipator
${\mathbb D}[\rho]$ typically drives the system to a noncritical
steady state, even when the Hamiltonian parameters are close to those
leading to a quantum transitions.  However, one may identify a regime
where the dissipation is sufficiently small to compete with the
coherent evolution driven by the critical Hamiltonian, leading to
potentially novel dynamic behaviors.  At such low-dissipation regime,
a dynamic scaling framework can be observed after appropriately
rescaling the global dissipation parameter $u$,
cf. Eq.~\eqref{lindblaseq}.  Indeed, the master Lindblad equation
(\ref{lindblaseq}) at the CQTs of the coherent Hamiltonian driving
develops a scaling behavior as well~\cite{NRV-19, RV-19, RV-20}, with
a further dependence on the dissipation scaling variable
\begin{equation}
  \gamma \equiv u / \Delta_L \,,
  \label{gammaupscaling}
\end{equation}
thus $\gamma\sim u L^{z}$.  Therefore, in the presence of dissipation
the dynamic scaling behavior~\eqref{scamag} after quenching $h$ is
expected to change into
\begin{equation}
  M(t,h_i,h,u,L) \approx L^{-\zeta} F_m(\theta,\kappa_i,\kappa,\gamma)\,.
  \label{scamagu}
\end{equation}

Our working hypothesis for the study of analogous issues at FOQTs is
that the dynamic scaling behavior~\eqref{scamagu} applies as well,
with the same definition~\eqref{gammaupscaling} of the dissipation
scaling variable $\gamma$.  In the following we challenge this
scenario by means of numerical computations.
For convenience, we calculate the rescaled longitudinal
magnetization, defined as
\begin{equation}
  \widetilde{M}(t,h_i,h,u,L)  = {M(t,h_i,h,u,L) \over  M(0,h_i,h,u,L) }\,,
  \label{rescmagdef}
\end{equation}
which is expected to behave as
\begin{equation}
  \widetilde{M}(t,h_i,h,u,L) 
  \approx \widetilde{F}_m(\theta,\kappa_i,\kappa,\gamma)\,,
  \label{scamagures}
\end{equation}
at both the CQT and FOQTs.  We recall that, according to our protocol,
the initial longitudinal magnetization at $t=0$ corresponds to that of
the equilibrium ground-state expectation value for $h_i$. Therefore
it satisfies the asymptotic scaling behavior~\cite{CPV-14, CNPV-14}
\begin{equation}
  M(0,h_i,h,u,L) \equiv M(h_i,L) \approx L^{-\zeta} f_m(\kappa_i)\,,
  \label{m0h}
\end{equation}
with $\zeta = 1/8$ at the CQT, $\zeta=0$ at the FOQTs, and $f_M$ being
a universal scaling function (apart from a multiplicative
normalization and a normalization of the argument) which depends on
the type of transition, being a CQT or a FOQT.

\section{Numerical results}
\label{foqtres}

In this section we present numerical results for the quantum Ising
chain subject to the protocol described in Sec.~\ref{model}, when the
system is close to the FOQT line, i.e.~for $g<1$ and $|h|\ll 1$.  We
also report some results at the CQT, for $g=1$ and $|h|\ll 1$,
extending the study already reported in Ref.~\cite{NRV-19}, which
focussed on the Kitaev fermionic wire (see also App.~\ref{kitaev}).
The latter is somehow related to quantum Ising chain, although they
are not equivalent, in particular in the presence of local
dissipation.  In all our simulations we set $\hslash =1$, and $J=1$ as
the energy scale.

Numerically solving the Lindblad master equation~\eqref{lindblaseq}
for a system as the one in Eq.~\eqref{hedef} generally requires a huge
computational effort, due to the large number of states in the
many-body Hilbert space $\mathcal{H}$, which increases exponentially
with the system size (dim $\mathcal{H} = 2^L$).  More precisely, the
time evolution of the density matrix $\rho(t)$, which belongs to the
space of the linear operators on $\mathcal{H}$, can be addressed by
manipulating a Liouvillian superoperator of size $2^{2L} \times
2^{2L}$. This severely limits the accessible system size to $L
\lesssim 10$ sites, unless the model is amenable to a direct
solvability.  A notable example in this respect is the Kitaev chain
with one-body Lindblad operators, whose corresponding Liouvillian
operator is quadratic in the fermionic creation and annihilation
operators (see, e.g., Ref.~\cite{NRV-19}).  Unfortunately this is not
the case for a dissipative Ising spin chain, in which the
Jordan-Wigner mapping of Lindblad operators as those in~\eqref{lxdef}
and~\eqref{lglobdef} produces a nonlocal string operator forbidding an
analytic treatment.  Therefore, in this work we resort to a
brute-force numerical integration of Eq.~\eqref{lindblaseq} through a
fourth-order Runge-Kutta method, with a time step $dt = 10^{-2}$,
sufficiently small to ensure convergence for all our purposes.

\subsection{Dynamics along the  FOQT line}
\label{numfoqt}

\begin{figure}[!t]
  \includegraphics[width=0.98\columnwidth]{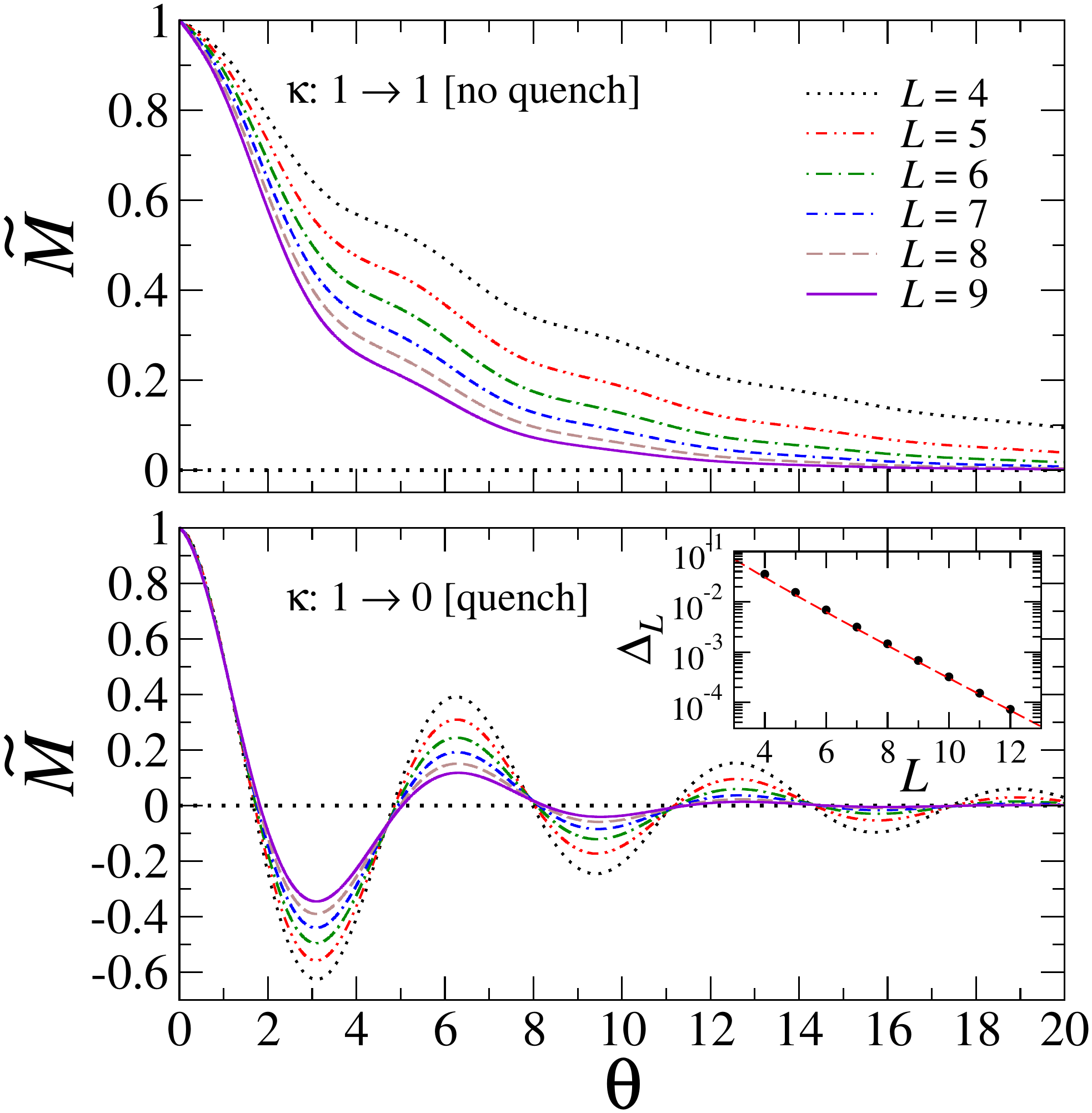}
  \caption{Time behavior of the longitudinal magnetization for a
    quantum Ising ring close to the FOQT, with $g=0.5$,
    in the presence of homogenous dissipation described by local
    Lindblad operators $\hat L^-_x$. We show the rescaled quantity
    $\widetilde{M}$ in Eq.~\eqref{rescmagdef} versus the rescaled
    time $\theta$, for different values of the system size $L$
    (see legend).  With increasing $L$, we keep the scaling
    variables $\kappa_i=1$ and $\gamma=0.1$ fixed.
    The upper panel is for $\kappa_i = \kappa$ (i.e., without the
    quench of the Hamiltonian parameter $h$), while
    the lower panel is for $\kappa = 0$.
    In the inset we show the energy gap $\Delta_L$
    as a function of $L$: black circles are the results
    obtained from the numerics, while the dashed red line
    denotes the estimate in Eq.~\eqref{deltapbc}.}
\label{fig:MagnLongT_FOQT_g05}
\end{figure}

In the following, we numerically challenge the dynamic scaling
behavior put forward in Sec.~\ref{dynbehsumm} at the FOQTs.  We
provide results for quantum Ising chains with $g=0.5$, up to lattice
sizes $L=O(10)$. Computations for other values of $g<1$ produce
analogous results and are not explicitly shown here.  Let us recall
that our dynamic protocol starts from the ground state of the
Hamiltonian~\eqref{hedef} with longitudinal-field parameter $h_i$;
then the system evolves according to the Lindblad master
equation~\eqref{lindblaseq}, with Hamiltonian parameter $h$ (in
principle different from $h_i$) and dissipative strength $u$.

Figure~\ref{fig:MagnLongT_FOQT_g05} displays the time evolution of the
longitudinal magnetization~\eqref{mtdef}, in particular the rescaled
one defined in Eq.~\eqref{rescmagdef}, for some selected values of
dynamic scaling variables.  The system-bath coupling has been modeled
through local dissipative mechanisms as in the top drawing of
Fig.~\ref{fig:sketch}, where the Lindblad operator associated to each
site induces incoherent lowering of the corresponding spin ($\hat
L_x^- = \hat \sigma_x^-$).
The scaling behavior~\eqref{scamagures} is checked by varying the
Hamiltonian and the dissipation parameters of the protocol with
increasing size $L$, so that the scaling variables $\kappa$,
$\kappa_i$, and $\gamma$ [as defined in Eqs.~\eqref{kappaising},
  \eqref{defkaika}, and \eqref{gammaupscaling}] are kept fixed.  For
the gap $\Delta_L$, entering the definition of the scaling variables
$\theta=\Delta_L t$ and $\gamma=u/\Delta_L$, we do not use its
asymptotic behavior~\eqref{deltapbc}, but the actual energy difference
of the lowest levels of the quantum Ising ring at $h=0$, with $L$
spins (it passes from $\Delta_L \approx 3.55 \times 10^{-2}$
for $L=4$, to $\Delta_L \approx 3.21 \times 10^{-4}$ for $L=10$,
as visible from the inset in the lower panel). 
In both cases considered in Fig.~\ref{fig:MagnLongT_FOQT_g05},
namely for $\kappa_i=\kappa$ (top panel, without quenching the
Hamiltonian) and for $\kappa_i \neq \kappa$ (bottom panel, in the
presence of a Hamiltonian quench), the longitudinal magnetization
appears to asymptotically vanish in the large-time limit, although
with different qualitative trends.  Actually this turns out to be a
general feature for any nonzero value of the dissipation variable
$\gamma$.  Analogous results are obtained using Lindblad operators
inducing incoherent raising of the corresponding spin ($\hat L_x^+ =
\hat \sigma_x^+$).

\begin{figure}[!t]
  \includegraphics[width=0.98\columnwidth]{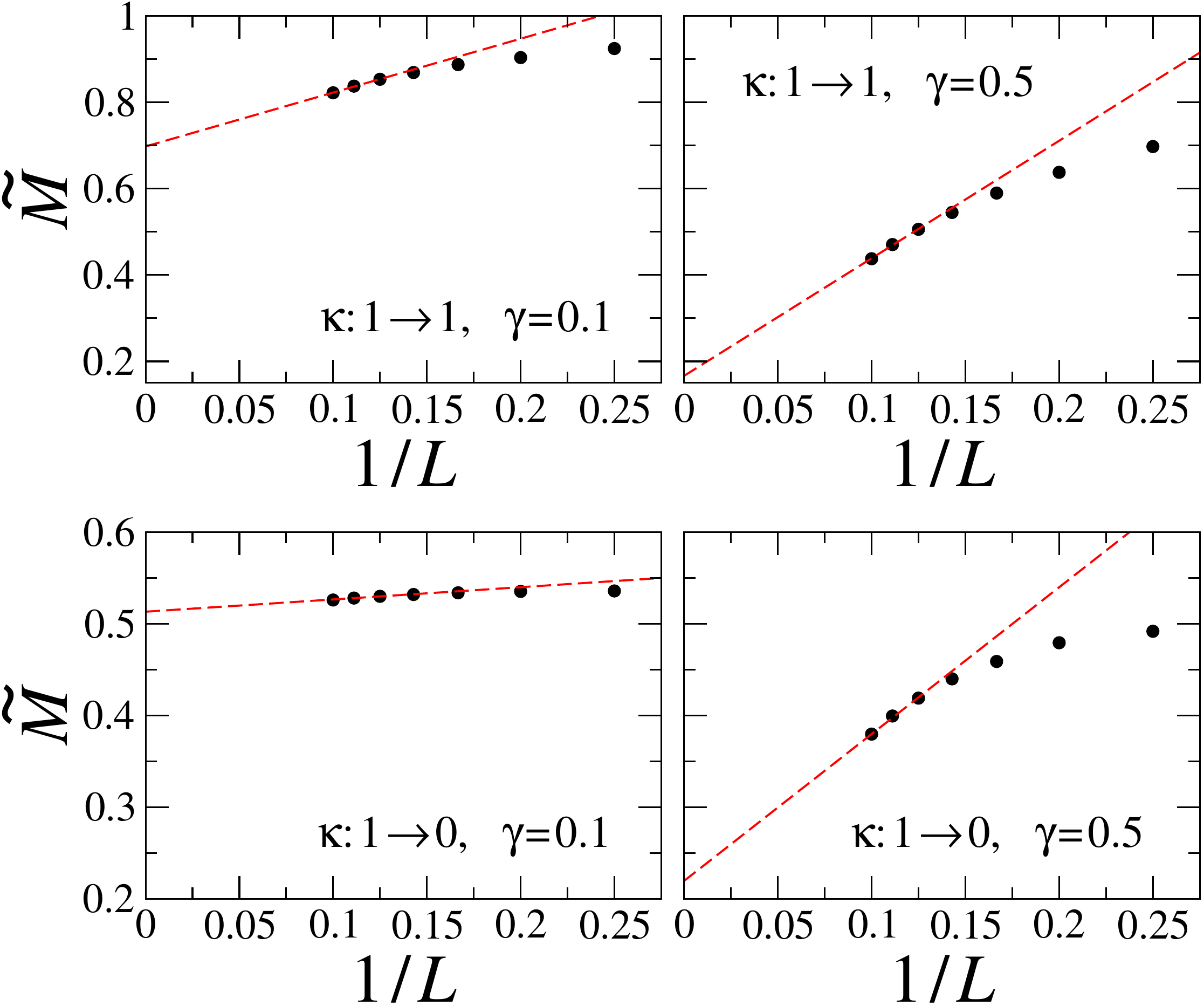}
  \caption{Approach to the asymptotic dynamic scaling behavior of the
    longitudinal magnetization at the FOQT, in the presence of
    dissipation $\hat L^-_x$.  Here we set $g=0.5$, $\kappa_i = 1$,
    and $\theta=1$, and show results for $\kappa=0,1$ and
    $\gamma=0.1$, $0.5$ (see indications in the panels).  The data are
    roughly compatible with a global $1/L$ approach to the asymptotic
    dynamic scaling, in particular for $\gamma=0.1$
    (the dashed red lines are drawn to guide the eye).}
\label{fig:approachtoscaling}
\end{figure}

Although limited to small system sizes, $L\leq 10$, our numerical
results substantially support the dynamic scaling behavior conjectured
in Eq.~\eqref{scamagures}, especially for sufficiently small values of
$\theta$.  Most likely, the large-$L$ convergence is not uniform and
tends to be slower with increasing $\theta$.  In particular, the
results reported in the lower panel of
Fig.~\ref{fig:MagnLongT_FOQT_g05} (when the dynamics arises both from
a quench of the Hamiltonian parameter and from the presence of
dissipation) display oscillations whose zeroes nicely scale with the
time scaling variable $\theta$, even for quite large $\theta$.  On the
other hand, the values at the maxima and minima undergo larger
corrections, likely requiring larger lattice sizes to clearly observe
the asymptotic scaling behavior.  The apparent asymptotic convergence
to the conjectured dynamic scaling is also suggested by the various
plots in Fig.~\ref{fig:approachtoscaling}, showing data at fixed
$\theta=1$, for various values of $\kappa$ and $\gamma$ (panels on the
left correspond to the two situations in
Fig.~\ref{fig:MagnLongT_FOQT_g05}, while panels on the right are for
the analogous cases with a larger dissipation strength $\gamma =
0.5$). The approach to the large-$L$ limit is generally compatible
with $1/L$ corrections, as hinted by the dashed red lines, denoting
$1/L$ fits of numerical data extrapolated at the largest available
sizes.

\subsection{Comparison with the single-spin problem}
\label{con1spin}

In the absence of dissipation, the dynamic scaling behavior of a
quantum Ising chain along the FOQT turns out to be well described
by an effective two-level problem~\cite{CNPV-14, PRV-18, NRV-19-wo, RV-20},
defined by the single-spin Hamiltonian
\begin{equation}
  \hat{H}_{s} = a_1 \, \hat\sigma^{(1)} + a_3 \,\hat\sigma^{(3)}.
  \label{ssham}
\end{equation}
Indeed the dynamic scaling functions of the quantum Ising chain, when
the boundary conditions do not favor any phase separated by the FOQT,
match the dynamics of the single-spin problem (see
App.~\ref{1spinModel} for an analytic discussion of the single-spin
problem). In a sense, closed systems behave rigidly at FOQTs.  We now
want to understand whether that kind of property persists in the
presence of dissipative interaction with an environment, such as in
the two schemes sketched in Fig.~\ref{fig:sketch}.

\begin{figure}[!t]
  \includegraphics[width=0.98\columnwidth]{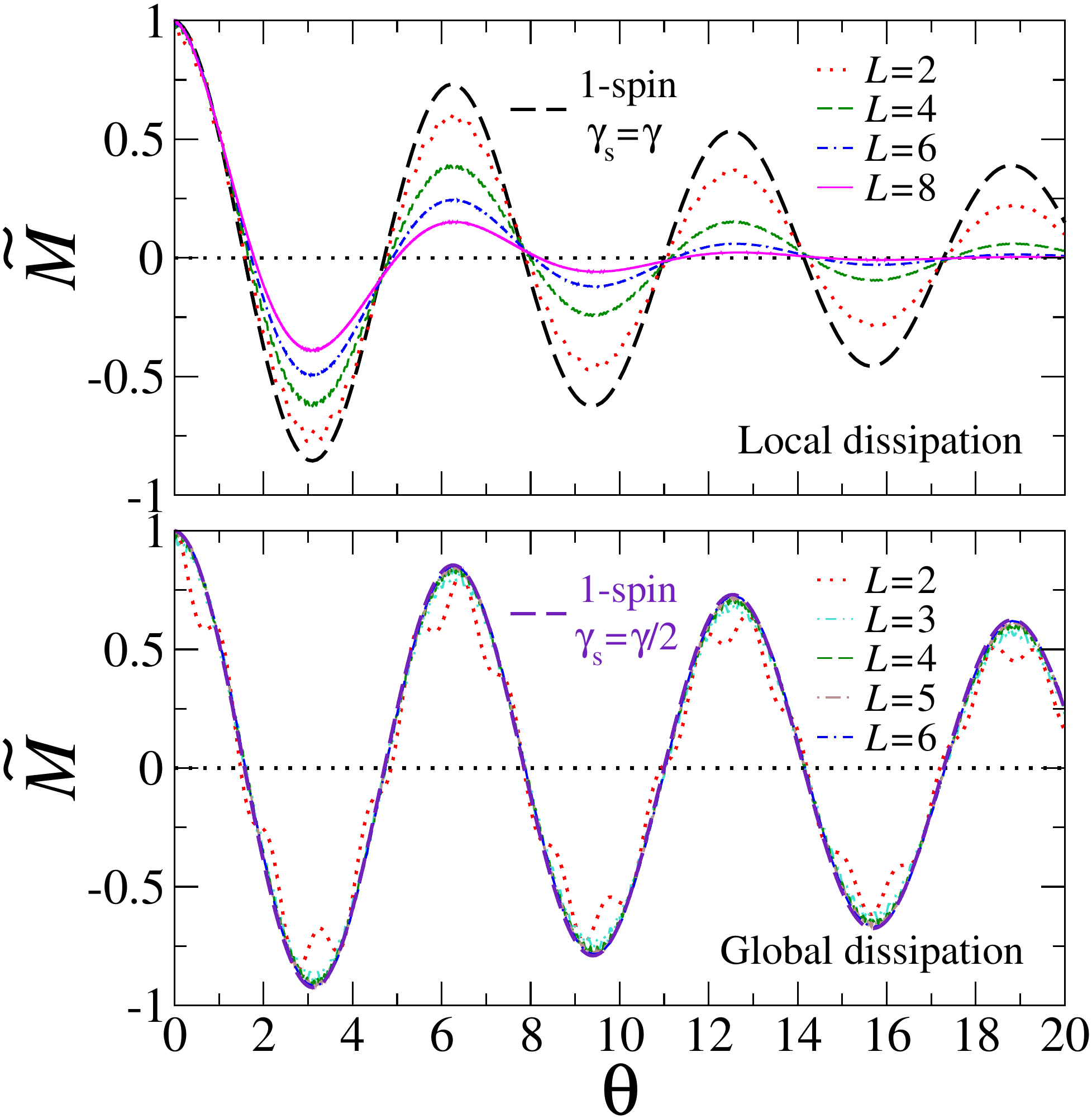}
  \caption{Time behavior of the longitudinal magnetization at $g=0.5$,
    in quench protocols with $\kappa_i=100$, $\kappa=0$, and
    $\gamma=0.1$, for various values of $L$ (see legends).  The system
    bath coupling is implemented in the form of either local
    dissipation operators~\eqref{dL}, with $\hat L_x = \hat
    \sigma^-_x$ (top panel), or a single global dissipation
    operator~\eqref{lglo}, with $\hat L = \Sigma^-$ (bottom panel).
    The numerical results are compared with the single-spin problem
    (thick-dashed curves): in the upper case, with increasing $L$, the
    curves tend to depart from the single-spin behavior; in the lower
    case, the curves appear to approach an asymptotic dynamic scaling
    behavior, which is well approximated by the solution of the
    single-spin problem with renormalized dissipation coupling
    $\gamma_s = \gamma/2 = 0.05$.  In both cases, the frequencies of
    oscillations reasonably match.}
\label{fig:dissvs1spin}
\end{figure}

In the presence of local dissipation (top drawing in
Fig.~\ref{fig:sketch}), we observe that the asymptotic scaling
behavior does not apparently display the rigidity property mentioned
above, in that the scaling functions are not reproduced by the
single-spin model.  The corresponding data for the magnetization as a
function of time are shown in the top panel of
Fig.~\ref{fig:dissvs1spin}. The initial oscillating behavior at finite
lengths is reasonably captured by the single-spin prediction with
dissipation coupling $\gamma_s = \gamma$ [thick-dashed black line ---
  see Eqs.~\eqref{a1theta}, \eqref{solkappa0}].  However, already for
$\theta \gtrsim 2$, the curves for $\widetilde{M}$ exhibit large
finite-size corrections and seem to approach an asymptotic overdamped
behavior for $L \to \infty$, quickly reaching the zero value. In any
case, the frequencies of oscillations for finite $L$ values match
those of the single-spin model, even for large $\theta$.

The rigidity of the dynamic scaling behavior can be recovered if a
global dissipative mechanism is considered (bottom drawing in
Fig.~\ref{fig:sketch}).  This is the case of the data reported in the
bottom panel of Fig.~\ref{fig:dissvs1spin}. The curves for different
$L$ appear to approach an asymptotic dynamic scaling behavior, as
well. Such convergence is much faster than that observed for the local
dissipation scheme.  Interestingly, the $L \to \infty$ behavior turns
out to be well approximated by the solution of the single-spin problem
(at least for $\theta\lesssim 5$), provided the dissipation coupling
is suitably renormalized according to $\gamma_s = \gamma/2$
(thick-dashed violet line).  We point out that this is a nontrivial
result, since we are matching the full many-body dynamics close to a
FOQT with a much simpler single-body behavior, by only admitting a
nonuniversal renormalization of the various scaling variables.

\subsection{Quantum work and heat}
\label{quthe}

We now discuss the quantum thermodynamics arising from the
out-of-equilibrium protocol.  As already mentioned in
Sec.~\ref{model}, a nonzero average work is only required at $t=0$,
when the Hamiltonian parameter suddenly changes from $h_i$ to $h$.
Therefore, it is the same as that of systems subject to quench
protocols without dissipation (i.e., for $u=0$).  The dynamic scaling
behavior of the work fluctuations for analogous quenches at the FOQTs
of the quantum Ising chain was studied in Ref.~\cite{NRV-19-wo}, showing
that they reproduce the analogous quantities of the single-spin
problem. In particular, the average work is given by
\begin{equation}
  W \approx \Delta_L\, {\cal W}(\kappa_i,\kappa) \,, \qquad
  {\cal W} = {(\kappa_i - \kappa)\kappa_i \over 2\sqrt{1+\kappa_i^2}}\,. 
  \label{wscal}
\end{equation}
Ref.~\cite{NRV-19-wo} also reports results for the higher moments of 
work fluctuations and a discussion of the correction to the
asymptotic behavior.

\begin{figure}[!t]
  \includegraphics[width=0.98\columnwidth]{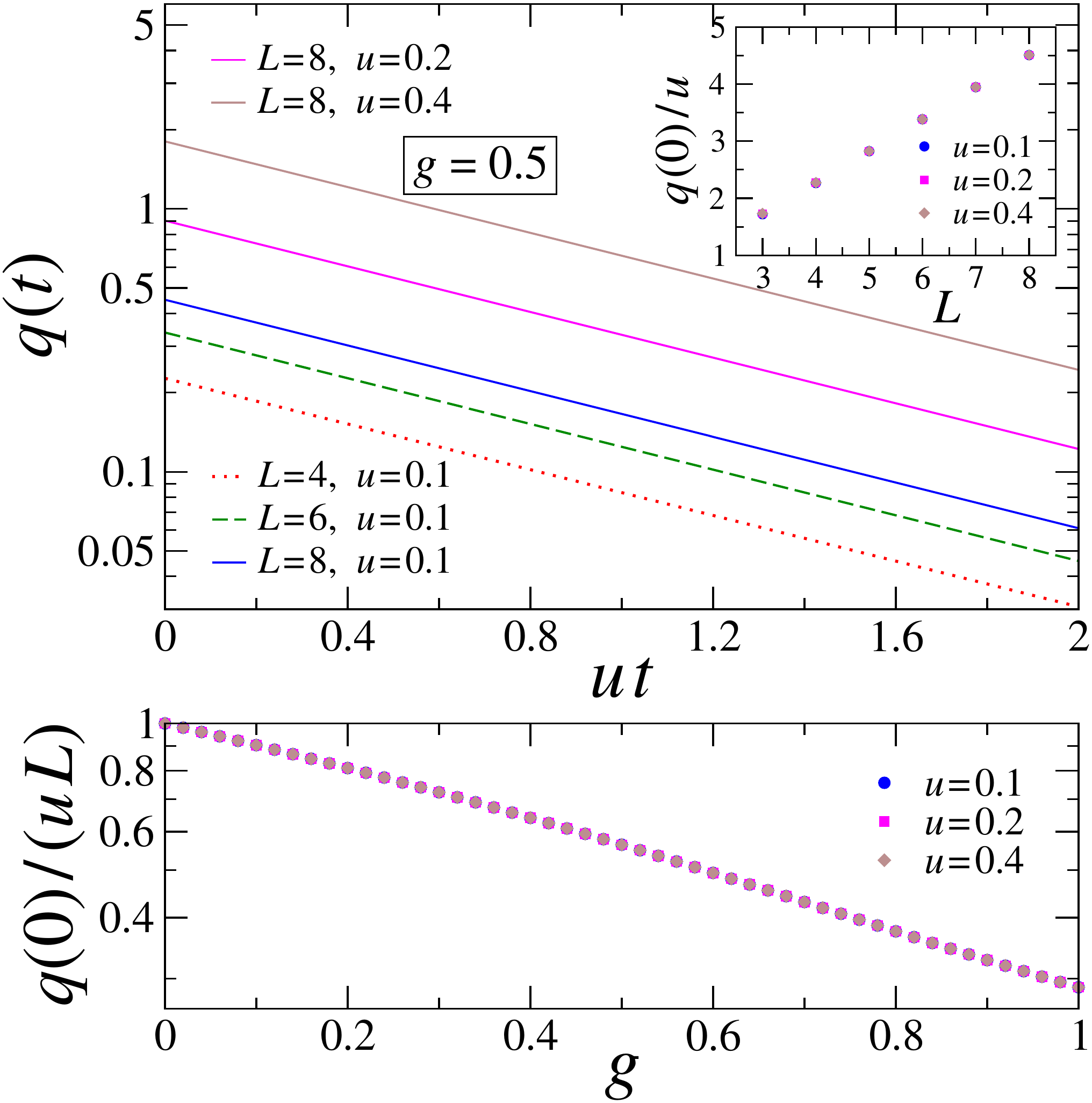}
  \caption{Behavior of the heat interchanged between the quantum Ising
    chain and the environment, per unit of time, as a function of
    the variables of the system. Top panel: the function $q(t)$
    with respect to the rescaled time $u\,t$, for different values
    of the system size $L$ and of the dissipation strength $u$ (see legend).
    In the inset we show the behavior of $q(t=0)/u$ with $L$,
    for various values of $u$. The transverse field strength is kept
    fixed at $g=0.5$.
    Bottom panel: the function $q(t=0) / (u\,L)$ with respect
    to $g$, for $L=6$ and various values of $u$.
    All the data reported in this figure as for $\kappa_i = \kappa=0$,
    while the system-bath coupling is taken in the form of local
    and uniform Lindblad operators $\hat L^-_x = \hat \sigma^-_x$.}
\label{fig:HeatRate_Lvar}
\end{figure}

Coming to the heat interchanged between the system and the environment,
for the sake of convenience here we focus on its time derivative
$q(t)=dQ/dt$, defined in Eq.~\eqref{aheat}.
Unlike for the average work, this quantity does not present any relevant
scaling property. In particular, our numerical results for the dynamics
of the quantum Ising chain homogeneously coupled to local incoherent
lowering operators, cf.~Eq.~\eqref{dL} with $\hat L^-_x = \hat \sigma^-_x$,
clearly indicate an exponential decay of $q(t)$ with time $t$,
as shown in the top panel of Fig.~\ref{fig:HeatRate_Lvar}.
More specifically, we found the following functional dependence
on the various parameters of the system:
\begin{equation}
  q(t) = C(g, \kappa_i, \kappa) \, u \, L \, e^{-u\,t} \,.
  \label{eq:qexp}
\end{equation}
The coefficient $C(g, \kappa_i, \kappa)$ depends nearly exponentially on
the transverse field $g$, although minor quantitative deviations are
present (bottom panel).  On the other hand, the dependence of $C$
on both $\kappa_i$ and $\kappa$ is very weak and tends to vanish
for increasing $L$ (on the scale of the figure,
it becomes unappreciable already for $L \gtrsim 8$).

Finally, it is worth emphasizing that the linear dependence of the rate
of exchanged heat $q(t)$ with the system volume ($q \propto L$,
as indicated in the top inset of Fig.~\ref{fig:HeatRate_Lvar})
hints at the fact that such quantity is essentially related to
a one-body mechanism, and not to a collective behavior of the system.
In fact, the exponential time dependence $q(t) \sim e^{- u t}$
is identical to the single-spin model behavior, as emerging from
the analysis in App.~\ref{1spinModel} [see, e.g., Eq.~\eqref{qranal}].

\subsection{Dynamics at the CQT}
\label{dyncqt}

\begin{figure}[!t]
  \includegraphics[width=0.98\columnwidth]{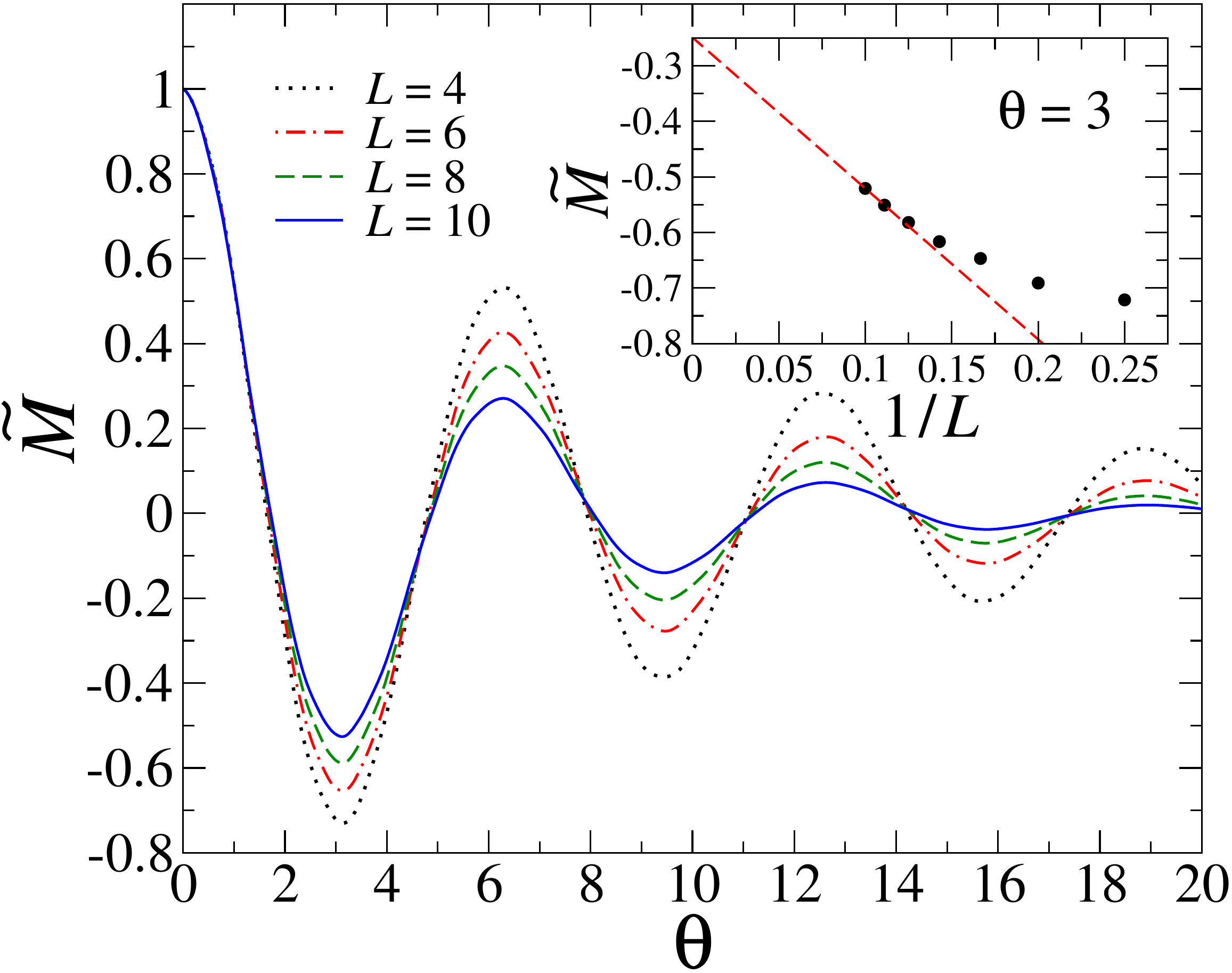}
  \caption{Time behavior of the longitudinal magnetization for a
    quantum Ising chain at the CQT, in the presence of homogenous
    dissipation described by local Lindblad operators $\hat L^-_x$. We
    show the rescaled magnetization $\widetilde{M}$,
    cf. Eq.~\eqref{rescmagdef}, versus the rescaled time $\theta$, for
    different values of the system size $L$ (see legend).  With
    increasing $L$ we keep the scaling variables $\kappa_i=1$,
    $\kappa=0$ and $\gamma=0.1$ fixed. The inset displays 
    data at fixed $\theta=3$ versus $1/L$, showing that 
    they are roughly compatible with a $1/L$ approach to the asymptotic
    dynamic scaling (the dashed red line is drawn to guide the eye),
    although further corrections are clearly visible.}
\label{fig:Magn_CQT_k1_ups01}
\end{figure}

We now discuss the dynamic scaling behavior at the CQT (i.e., at $g=1$
and $|h|\ll 1$), focussing on the longitudinal magnetization [see
  Eq.~\eqref{scamagures}]. Figure~\ref{fig:Magn_CQT_k1_ups01} reports
the rescaled quantity $\widetilde{M}$ versus the rescaled time
$\theta= t L^{-z}$, for various systems sizes, up to $L=10$.  The
curves with increasing $L$ are obtained keeping the scaling variables
$\kappa_i\sim h_i L^{y_h}$, $\kappa \sim h L^{y_h}$, and $\gamma \sim
u L^z$ (with $y_h=15/8$ and $z=1$) fixed.  The displayed data are for
$\kappa_i=1$, $\kappa=0$, $\gamma = 0.1$, but analogous qualitative
conclusions can be drawn by changing the specific values of such
rescaled quantities.
Our results substantially support the dynamic scaling behavior
conjectured in Eq.~\eqref{scamagures}, in particular for sufficiently
small values of $\theta$. The approach to the asymptotic behavior is
again compatible with $1/L$ corrections, as expected (see the inset,
where numerical data for the rescaled magnetization at fixed
$\theta=3$ are plotted against the inverse system size, and the dashed
red line denotes a $1/L$ fit of such data at large $L$).
We point out that analogous dynamic scaling behaviors have been
reported for the Kitaev fermionic wire (see App.~\ref{kitaev}), where
much larger sizes can be reached, allowing us to achieve a definitely
more robust check of the dynamic scaling behavior
at a CQT~\cite{NRV-19, RV-19}.

Again the quantum thermodynamics arising from the dynamic protocol is
characterized by an initial work at $t=0$, due to the quench of the
Hamiltonian parameter. Its scaling behavior was already discussed in
Ref.~\cite{NRV-19-wo} for closed systems.  As demonstrated there, the
average work shows the scaling behavior
\begin{equation}
  W \approx L^{-z} \, {\cal W}(\kappa_i,\kappa)\,.
  \label{wscalcqt}
\end{equation}
In App.~\ref{kitaev} we discuss the average work within the Kitaev
fermionic wire at its CQT, where the quench is performed over the
chemical potential, corresponding to the transverse field $g$ of the
Ising chain. As shown there, the leading contribution to the average
work is provided by analytical terms, while the scaling part, such as
that in Eq.~\eqref{wscalcqt}, turns out to be subleading.

On the other hand, even at the CQT, the heat interchanged with
the environment does not exhibit scaling properties.
We have observed an exponential decay in time and a trivial
linear dependence with $L$, for the function $q(t)$,
analogously as in Eq.~\eqref{eq:qexp}.
The only difference with respect to the $g<1$ case is an
enhanced sensitivity of the coefficient $C(g=1, \kappa_i, \kappa)$
on the rescaled longitudinal field parameters, $\kappa_i$ and $\kappa$,
which however rapidly reduces with increasing $L$.

\section{Conclusions}
\label{conclu}

We have investigated the effects of dissipation on the quantum
dynamics of many-body systems close to a FOQT (that is, whose
Hamiltonian parameters are those leading to a FOQT for the closed
system), arising from the interaction with the environment, as for
example sketched in Fig.~\ref{fig:sketch}. The latter is modeled
through a class of dissipative mechanisms that can be effectively
described by Lindblad equations~\eqref{lindblaseq} for the density
matrix of the system~\cite{BP-book, RH-book}, with local or global
homogenous Lindblad operators, such as those reported in
Eqs.~\eqref{dL}-\eqref{lxdef} or~\eqref{lglo}-\eqref{lglobdef},
respectively.  This framework is of experimental interest, indeed the
conditions for its validity are typically realized in quantum optical
implementations~\cite{SBD-16}.  We have analyzed how homogenous
dissipative mechanisms change the dynamic scaling laws developed by
closed systems at FOQTs (see, e.g., Refs.~\cite{PRV-18, PRV-20}).  We
also mention that analogous issues have been addressed at CQTs (see,
e.g., Refs.~\cite{NRV-19, RV-19, RV-20}).

We study the above issues within the paradigmatic one-dimensional
quantum Ising model, cf.~Eq.~\eqref{hedef}, which provides an optimal
theoretical laboratory for the investigation of phenomena emerging at
quantum transitions.  Indeed its zero-temperature phase diagram
presents a line of FOQTs driven by a longitudinal external field,
ending at a continuous quantum transition. To investigate the
interplay between coherent and dissipative drivings at the
quantum-transition line (FOQTs for $g<1$ and $|h|\ll 1$ and CQT for
$g\approx 1$ and $|h|\ll 1$), we consider the following dynamic
protocol.  The system is initialized, at $t=0$, into the ground state
of the Hamiltonian~\eqref{hedef}, for a given longitudinal parameter
$h_i$; then, for $t>0$, it evolves according to the Lindblad
equation~\eqref{lindblaseq}, where the coherent driving is provided by
$\hat{H}(g,h)$, with $h$ generally different from $h_i$, and the
system-bath interaction effectively described by the dissipator
${\mathbb D}[\rho]$ with a fixed coupling strength $u$.

Analogously to what happens at CQTs, we observe a regime where the
system develops a nontrivial dynamic scaling behavior, which is
realized when the dissipation parameter $u$ scales as the energy
difference $\Delta$ of the lowest levels of the Hamiltonian of the
many-body system.  However, unlike CQTs where $\Delta$ is power-law
suppressed, at FOQTs $\Delta$ is exponentially suppressed when
boundary conditions do not favor any particular phase~\cite{CNPV-14,
  PRV-20}.  Numerical solutions of the Lindblad equations up to
lattice sizes with $L\approx 10$ substantially confirm the existence
of such dynamic scaling behavior and pave the way toward experimental
testing in the near future through quantum simulation platforms for
spin systems of small size, where the required resources are less
demanding.

We also compare the emerging asymptotic scaling behavior with the
single-spin problem interacting with an environment modeled by a
corresponding Lindblad equation.  Unlike closed systems where the
unitary dynamics is well described by the two-level single-spin
problem, in the presence of dissipation the system loses such a
rigidity, and the scaling behavior turns out to significantly differ.
These changes arise from the competition of the rigidity
properties of the system at the FOQT and the local dissipation that
tends to destroy the nonlocal rigidity. On the other hand, in the case
of global dissipators the dynamic scaling resembles that of the
single-spin problem.  Since they treat globally the system, it is not
surprising that the resulting dynamics maintains the main features of
the single-spin scenario.

The arguments leading to this scaling scenario at FOQTs are quite
general.  Analogous phenomena are expected to develop in any
homogeneous $d$-dimensional many-body system at a continuous quantum
transition, whose Markovian interactions with the bath can be
described by local or extended dissipators within a Lindblad
equation~\eqref{lindblaseq}.

\appendix

\section{The single-spin model}
\label{1spinModel}

We consider the following single-spin Hamiltonian:
\begin{equation}
\hat{H}_s = \Delta \, \hat{H}_r \,,\qquad \hat{H}_r = {1\over 2}
\,\hat\sigma^{(3)} - {\kappa\over 2} \, \hat\sigma^{(1)}\,,
\label{Hsig}
\end{equation}
where $\Delta$ is the energy scale (the gap for $\kappa=0$), $\kappa$
is the rescaled parameter related to the intensity of an applied
external magnetic field, and $\hat\sigma^{(k)}$ are the usual spin-1/2
Pauli matrices.
We discuss the dynamics of that single-spin system
subject to dissipation, as described by the Lindblad master equation
\begin{eqnarray}
&&  {\partial\rho\over \partial t} = -{i\over \hslash} \big[ \hat
    H_s,\rho \big] + u \,{\mathbb D}(\rho)\,,
  \label{lindblaseqs}\\
&&{\mathbb D}(\rho) = 
  \hat L \rho \hat L^\dagger - \tfrac{1}{2}
  \big( \rho\, \hat L^\dagger \hat L + \hat L^\dagger \hat L \rho \big)\,.
\end{eqnarray}

The protocol starts again from the ground state associated with an
initial value $\kappa_i$, that is, the pure state
\begin{equation}
  |\Psi(t=0)\rangle = \cos(\alpha_i/2) \, |+\rangle + 
  \sin(\alpha_i/2) \, |-\rangle \,,
  \label{eigstatekappai}
\end{equation}
where $|\pm \rangle$ are the eigenstates of $\hat\sigma^{(1)}$ and
$\tan (\alpha_i) = \kappa_i^{-1}$.  The corresponding density matrix
reads
\begin{align}
  &\rho(t=0) = {1\over 2} \left[ \mathbb{\hat I} + c_k \hat\sigma^{(k)}\right]\,,
  \label{rhoin}\\
  & c_1 = \cos \alpha_i\,, \quad c_2= 0 \,, \quad c_3 = \sin \alpha_i \,,
  \nonumber 
\end{align}
where $\mathbb{\hat I}$ is the $2 \times 2$ identity operator.
Then the evolution is determined by the Lindblad
equation~\eqref{lindblaseqs} with coherent driving given by the
Hamiltonian at $\kappa$ (thus for $\kappa\neq \kappa_i$ we have also a
quench).

We may generally define the energy of the system as
\begin{equation}
  E = {\rm Tr} \big[ \rho \hat{H}_s \big] \,,
  \label{energy}
\end{equation}
whose time derivative allows us to define quantities analogous to the
heat $q$ and work $w$ in the time unit, i.e.
\begin{equation}
  {dE\over dt} = {\rm Tr} \left[{d\rho\over dt} \hat{H}_s \right]+
  {\rm Tr} \bigg[ \rho {d\hat{H}_s\over dt} \bigg] \equiv q + w\,.
\label{timederene}
\end{equation}
Using the Lindblad master equation we can easily derive the relation
\begin{equation}
  q = {d Q\over dt} = {\rm Tr} \left[ {d\rho\over dt} \hat{H}_s \right] =
  u \, {\rm Tr} \big[ {\mathbb D}(\rho) \hat{H}_s \big] \,.
  \label{qeq}
\end{equation}
One can easily prove that $q=0$, if $[\hat H, \hat L] = 0$.

Moreover we define the {\em purity}
\begin{equation}
  P = {\rm Tr} \big[ \rho^2 \big]\,, 
  \label{purity}
\end{equation}
which equals one for pure systems.  Its time derivative can be written
as
\begin{equation}
  {dP\over dt} = 2 \, {\rm Tr}\left[ {d\rho\over dt} \rho\right] = 2 u 
  {\rm Tr} \big[ {\mathbb D}(\rho)\rho \big]\,.
  \label{dpeq}
\end{equation}

We may rescale the parameters and the time variable so that
\begin{align}
&\rho^\prime(\theta)
\equiv  {\partial\rho\over \partial \theta} = -{i\over \hslash} \big[ 
    \hat{H}_r,\rho \big] + \gamma \,{\mathbb D}(\rho)\,,
  \label{lindblaseqr}\\
&\theta = t\Delta\,,\qquad \gamma = u/\Delta\,.\label{rescvar}
\end{align}
The time dependent density matrix can be generally parametrized as
\begin{equation}
  \rho(\theta) = {1\over 2} \left[  \mathbb{\hat I} + A_k(\theta) \hat \sigma^{(k)} \right] \,,
  \qquad \sum_k A_k(\theta)^2 \le 1 \,,
  \label{rhot}
\end{equation}
where $A_i$ are real functions of the rescaled time $\theta$.  Note that
\begin{equation}
  {\rm Tr} [\rho] = 1\,,\qquad {\rm Tr} \big[ \rho^2 \big] = 
  {1\over 2} \Big( 1 + \sum_k A_k^2 \Big) \,.
  \label{rhotr}
\end{equation}

The Lindblad equation~\eqref{lindblaseqs} can be turned into coupled
differential equations of the functions $A_k$.
Let us consider the Lindblad operators
\begin{equation} 
  \hat L^{\pm}\equiv \hat\sigma^{\pm} \equiv {\hat\sigma^{(1)} \pm i 
    \hat\sigma^{(2)}\over 2}\,,
  \label{ldef}
\end{equation}
corresponding to the sign $+$ and $-$ respectively.  Straightforward
computations lead to the coupled differential equations
\begin{eqnarray}
  && A_1^\prime = - A_2 - {\gamma\over 2} A_1  \,, \nonumber \\
  && A_2^\prime = A_1 + \kappa A_3 - {\gamma\over 2} A_2   \,, \nonumber \\
  && A_3^\prime = - \kappa A_2  - \gamma (A_3\mp 1)  \,. \label{diffeq}
\end{eqnarray}
The upper/lower signs correspond to the cases $\hat L^{\pm}$.  The
various observables can be written in terms of the function $A_i$.
For example, the longitudinal magnetization $M$ reads
\begin{equation}
  {\cal M}(\theta) = {\rm Tr} \big[ \hat\sigma^{(1)} \rho(\theta) \big]
  = A_1(\theta) \,.
\label{a1theta}
\end{equation}
The heat per unit of rescaled time is given by
\begin{eqnarray}
  q_r &\equiv& Q^\prime = {\rm Tr} \big[ \rho^\prime \hat{H}_r \big]=
  {1\over 2} A_3^\prime  - {\kappa\over 2} A_1^\prime 
  \nonumber \\
  &=& - {\gamma\over 2} \left(  A_3\mp 1\right)   + {1\over 4} \gamma\kappa A_1  \,.
  \label{qrres}
\end{eqnarray} 
The time dependence of the purity~\eqref{purity} can be easily derived
using Eq.~\eqref{rhotr} and the above solutions, obtaining
\begin{eqnarray}
  P(\theta) &=& {1\over 2} \Big( 1  + \sum_k A_k(\theta)^2 \Big) \le 1\,,
  \label{psol}\\
  P^\prime(\theta) &=&\sum_k A_k(\theta) A_k^\prime(\theta) = \nonumber\\
  &=&- {\gamma\over 2} (A_1^2 + A_2^2) - \gamma A_3 (A_3\mp 1)\,.
  \label{dpsol}
\end{eqnarray}

In the case $\kappa=0$, one can easily find the solution
\begin{eqnarray}
  && A_1(\theta) = e^{-\gamma \theta/2} 
  \left[A_1(0) \, {\rm cos}(\theta) - A_2(0) \, {\rm sin}(\theta) \right]
  \,,\qquad
  \label{solkappa0}\\
  && A_2(\theta) = e^{-\gamma \theta/2} 
  \left[A_1(0) \, {\rm sin}(\theta) + A_2(0) \, {\rm cos}(\theta) \right]
  \,,\nonumber\\
  && A_3(\theta)\mp 1 = e^{-\gamma \theta} [A_3(0)\mp 1]
  \,,\nonumber
\end{eqnarray}
in terms of the initial density matrix $\rho(0)$, and in particular of
its coefficients $A_i(0)$.  Therefore we have that
\begin{eqnarray}
  q_r & = & 
  - {\gamma\over 2} e^{-\gamma \theta} [A_3(0)\mp 1]\,,
  \label{qranal}\\
  Q & = & \Delta \int_0^\infty d\theta \, q_r = - {\Delta \over 2}[A_3(0)\mp 1]\,.
  \label{Qtot}
\end{eqnarray}
Note that $Q$ is positive/negative for pumping/decay (positive $Q$
means that the system is getting energy from the bath).  The
purity~\eqref{purity} turns out to exponentially approach one,
reflecting the fact the the system relaxes to a pure state.

In the case of {\em dephasing} Lindblad operator
\begin{equation}
  \hat L_d \equiv \hat \sigma^{(3)}\,,
  \label{deplin}
\end{equation}
we obtain the coupled differential equations
\begin{eqnarray}
  A_1^\prime & = & - A_2 - 2\gamma A_1  \,, \nonumber \\
  A_2^\prime & = & A_1 + \kappa A_3 - 2\gamma A_2   \,, \nonumber \\
  A_3^\prime & = & -\kappa A_2\,. \label{diffeqd}
\end{eqnarray}
Using the above equations, we derive the heat per unit of rescaled
time, which is given by
\begin{equation}
  q_r  = {\rm Tr} [\rho^\prime \hat{H}_r]=  \gamma\kappa A_1\,.
  \label{qrres2}
\end{equation}

Again, for $\kappa=0$ the solution is quite simple, obtaining
\begin{eqnarray}
  A_1(\theta) & = & e^{-2\gamma \theta} 
  \left[A_1(0) \, {\rm cos}(\theta) - A_2(0) \, {\rm sin}(\theta) \right]
  \,,\qquad \nonumber \\
  A_2(\theta) & = & e^{-2\gamma \theta} 
  \left[A_1(0) \, {\rm sin}(\theta) + A_2(0) \, {\rm cos}(\theta) \right]
  \,,\nonumber\\
  A_3(\theta) & = & A_3(0)\,.
  \label{solkappa0d}
\end{eqnarray}
No heat transmission occurs, because $[\hat H, \hat L] = 0$ when
$\kappa=0$.  On the other hand the purity changes
\begin{equation}
  P(\theta) = {1\over 2} \Big\{ 1 + A_3(0)^2 + e^{-4\gamma\theta}
  \big[ A_1(0)^2 + A_2(0)^2 \big] \Big\}\,.
  \label{ptausig3}
\end{equation}
approaching exponentially the asymptotic value
\begin{equation}
  P(\theta\to\infty) = {1 + A_3(0)^2\over 2}\,.
  \label{asypval}
\end{equation}
Therefore, the spin system relaxes to a mixed state under dephasing.

\section{Quantum thermodynamics of a fermionic wire coupled to local baths}
\label{kitaev}

In this appendix we focus on the quantum thermodynamics of fermionic
quantum wires coupled to local Markovian baths.  In particular, we
consider a Kitaev quantum wire defined by the
Hamiltonian~\cite{Kitaev-01}
\begin{equation}
  \hat H_K = - J \sum_{x=1}^L \big( \hat c_x^\dagger \hat
  c_{x+1} + \delta \, \hat c_x^\dagger \hat c_{x+1}^\dagger+{\rm h.c.}
  \big) - \mu \sum_{x=1}^L \hat n_x \,,
  \label{kitaev2}
\end{equation}
where $\hat c_x$ is the fermionic annihilation operator associated
with the sites of the chain of size $L$, $\hat n_x\equiv \hat
c_x^\dagger \hat c_x$ is the density operator, and $\delta>0$.  We set
$\hslash =1$, and $J=1$ as the energy scale.  We consider antiperiodic
boundary conditions, $\hat c_{L+1} = - \hat c_1$, and even $L$ for
computational convenience.

The Hamiltonian~\eqref{kitaev2} can be mapped into a spin-1/2 XY
chain, through a Jordan-Wigner transformation~\cite{Sachdev-book}.  In
the following we fix $\delta=1$ (without loss of generality), so that
the corresponding spin model is the quantum Ising chain~\eqref{hedef}.
Note however that the non-local Jordan-Wigner transformation of the
Ising chain with periodic or antiperiodic boundary conditions does not
map into the fermionic model~\eqref{kitaev2} with periodic or
antiperiodic boundary conditions. Indeed further considerations
apply~\cite{Katsura-62, Pfeuty-70}, leading to a less straightforward
correspondence, depending on the parity of the particle number
eigenvalue.  Therefore, the Kitaev quantum wire cannot be considered
completely equivalent to the quantum Ising chain (see, e.g., the
discussion in the appendix of Ref.~\cite{RV-20}). However, analogously
to the quantum Ising chain, the Kitaev quantum wire undergoes a
continuous quantum transition at $\mu=\mu_c = -2$, between a
disordered ($\mu<\mu_c$) and an ordered quantum phase
($|\mu|<|\mu_c|$).  This transition belongs to the two-dimensional
Ising universality class~\cite{Sachdev-book}, characterized by the
length-scale critical exponent $\nu=1$, related to the
renormalization-group dimension $y_\mu = 1/\nu=1$ of the Hamiltonian
parameter $\mu$ (more precisely of the difference $\bar{\mu} \equiv
\mu-\mu_c$).  The dynamic exponent associated with the unitary quantum
dynamics is $z=1$.

We focus on the out-of-equilibrium thermodynamic behavior of the Fermi
lattice gas close to its continuous quantum transition in the presence
of homogeneous dissipation mechanisms described by the Lindblad
equation~\eqref{lindblaseq}.  We consider local dissipative
mechanisms, so that ${\mathbb D}[\rho] = \sum_x {\mathbb D}_x[\rho]$
is given by a sum of local (single-site) terms (top drawing of
Fig.~\ref{fig:sketch}). The onsite Lindblad operators $\hat L_x$
describe the coupling of each site with an independent bath ${\mathcal
  B}$, associated with particle loss (l) or pumping (p), thus
\begin{equation}
  \hat L_{{\rm l},x} = \hat c_x, \qquad \hat L_{{\rm p},x} = \hat c_x^\dagger\,,
  \label{liopki}
\end{equation}
respectively.  With this choice of dissipators, the full open-system
many-body fermionic master equation enjoys a particularly simple treatment.

As a matter of fact, we mention that the out-of-equilibrium dynamics
of the Kitaev quantum chain with such kind of dissipators has been
object of several studies in different contexts
(see, e.g., Refs.~\cite{Prosen-08, HCG-13, KMSFR-17, BBD-20}).
One of them concerns the behavior of the fermionic correlation
functions resulting from a sudden quench~\cite{NRV-19, RV-19}.
Namely, one considers a protocol that starts from the ground state
$|0_{\mu_i}\rangle$ of $\hat H_K$ for a generic $\mu_i$.
The system is then let free to evolve after a
quench of the Hamiltonian parameter to $\mu$, at $t=0$ and a
simultaneous turning on of the interaction with the environment
controlled by the dissipation coupling $u$. 
Below we discuss further aspects of such dynamics,
which have not been considered so far.

\begin{figure}[!t]
  \includegraphics[width=0.98\columnwidth]{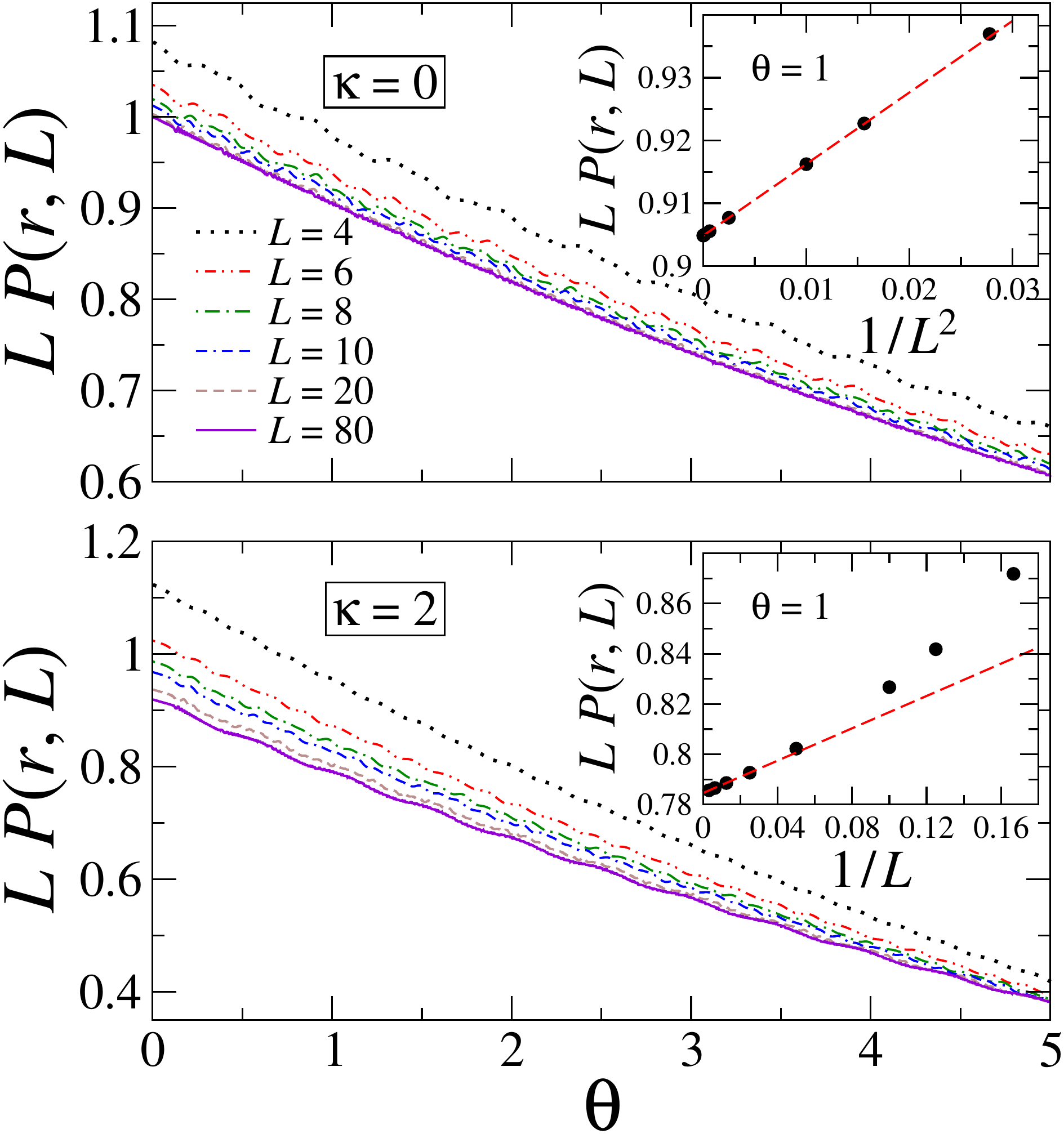}
  \caption{Time behavior of the correlation function $P(r,t)$ for the
    Kitaev quantum wire close to the CQT point ($\delta=1$, $\mu =
    -2$).  The system is driven out of equilibrium by the coupling
    with an external bath in the Lindblad form, under the form of
    local and uniform incoherent particle losses with rescaled
    strength $\gamma=0.1$.  The top figure shows data for $\kappa_i =
    \kappa = 0$ (i.e., the Hamiltonian is exactly at the CQT), while
    the bottom figure is for $\kappa_i = \kappa = 2$.  The various
    curves in the two main panels show $L\, P(r,t)$ for $r=L/2$ and
    different chain lengths $L$ (see legend), as a function of the
    rescaled time $\theta$.  In the insets we report the same quantity
    as a function of $L^{-2}$ (top) or of $L^{-1}$ (bottom), for a
    fixed value of $\theta = 1$.  To perform this analysis, we have
    first washed out the wiggles in $\theta$ (see main plot), by
    fitting the original curves in the interval $\theta \in [0,5]$
    with a fifth-order polynomial.}
\label{fig:Kitaev_P_ups01}
\end{figure}

Before presenting the analysis of the quantum thermodynamic properties
during the time evolution associated with the above protocol,
let us come back to the following correlation function at distance $r$:
\begin{equation}
  P(r,t) = {\rm Tr} \big[ \rho(t)\,(\hat c_x^\dagger \hat c_{x+r}^\dagger +
    \hat c_{x+r} \hat c_{x}) \big] \,. \label{ptf}
\end{equation}
This is expected to converge, in the large-$L$ limit, to the asymptotic
dynamic scaling behavior~\cite{NRV-19}
\begin{equation}
  P(r,t,\bar{\mu}_i,\bar{\mu}_f,u,L) \approx 
  L^{-1} {\cal P}(R,\theta,\kappa_i,\kappa,\gamma)\,,
  \label{dFSSd}
\end{equation}
where $R=r/L$ and the other scaling variables are defined as usual.
The data reported in Fig.~\ref{fig:Kitaev_P_ups01} display the
approach to such scaling behavior, after keeping the scaling variables
fixed, where
\begin{eqnarray}
  &\theta = t L^{-z} \,, \quad & \gamma = u L^z\,,\qquad
  \kappa(\mu) = \bar{\mu} L^{y_\mu}\,,
  \label{dynscavarkii}
  \\
  &\bar{\mu}  = \mu - \mu_c \,,\quad
  &\kappa_i \equiv \kappa(\mu_i)\,,\qquad
  \kappa\equiv \kappa(\mu)\,.
  \nonumber 
\end{eqnarray}
The large-$L$ asymptotic behavior turns out to be approached with
power-law suppressed corrections, analogously to what has been hinted
in the main text for the dissipative quantum Ising chain.  The
approach to the asymptotic behavior is generally characterized by
$O(1/L)$ corrections, except at the critical point where they may get
suppressed by a larger $O(1/L^2)$ power~\cite{CPV-14,RV-19,RV-20}, as also
shown by the insets of Fig.~\ref{fig:Kitaev_P_ups01}.  Further results
for the asymptotic large-$L$ behavior of the fermionic correlations
functions can be found in Refs.~\cite{NRV-19, RV-19}.

We now extend the analysis of the out-of-equilibrium dynamics arising
from the protocol described above to the quantum thermodynamics,
focussing on the quantum work and heat interchange during the
out-of-equilibrium evolution. We discuss the case of particle decay as
a dissipative mechanism; the case of pumping can be easily obtained by
analogous computations.

The first law of thermodynamics~\eqref{filaw} describes the energy
flows of the global system, including the environment.  In our
protocol, a nonvanishing work is only done at $t=0$, if the
Hamiltonian parameter $\mu$ is suddenly changed from $\mu_i$ to
$\mu\neq \mu_i$. Since after quenching the Hamiltonian is kept fixed,
thus $w(t)=0$ for $t>0$, and the average work is just given by
\begin{eqnarray}
  W & = & \langle 0_{\mu_i} | \hat H_K(\mu) - \hat H_K(\mu_i) |
  0_{\mu_i}\rangle \nonumber\\ &=& (\mu_i - \mu ) L\, \langle 0_{\mu_i}
  | \hat n_x | 0_{\mu_i}\rangle \,,
  \label{woki}
\end{eqnarray}
where last expression is obtained because the antiperiodic boundary
conditions respect translational invariance.  The matrix element of
the particle density $\hat n_x$ can be computed analytically, in
particular when $\mu_i=\mu_c$, we have
\begin{align}
&  \langle 0_{\mu_c} | \hat n_x | 0_{\mu_c}\rangle = D(L)\,,
  \label{omucn} \\
&  D(L)= {1\over 2} \! - \!
  {\sin\left({\pi\over 2L}\right) \over L \left[1 -
      \cos\left({\pi\over L}\right)\right]}
  = {\pi-2\over 2\pi} \! + \!{\pi\over 24 L^2} \!+ \! O(L^{-4})\,.
  \label{omucnex}
\end{align}
More generally, for $\mu$ in the critical region, so that
$\bar{\mu}\equiv \mu-\mu_c \ll 1$, we obtain the asymptotic
expansion~\cite{CPV-14}
\begin{equation}
  \langle 0_{\mu} | \hat n_x | 0_{\mu}\rangle = 
  f_a(\bar\mu) + L^{-\zeta} f_s(\kappa) + O(L^{-2})\,,
  \label{omun}
\end{equation}
where $\zeta = 1 + z - y_\mu = 1$, $f_a$
and $f_s$ are appropriate functions.

Concerning the heat interchanged with the environment, we derive the
nontrivial relation
\begin{equation}
  q(t) = - u \, {\rm Tr} \big[ \rho(t) \hat H_K(\mu) \big] \,,
\label{qrel}
\end{equation}
obtained by replacing $\partial_t \rho(t)$ using the Lindblad equation
for the density matrix, and further manipulations related to the
particular structure of the Lindblad dissipator ${\mathbb D}_j[\rho]$.
Moreover, since the Hamiltonian is independent of the time for $t>0$,
we also have
\begin{equation}
  q(t) = {\rm Tr} \bigg[ {d\rho(t)\over dt} \hat H_K(\mu) \bigg]
  ={d\over dt} {\rm Tr} \big[ \rho(t) \hat H_K(\mu) \big] = {d E_s\over dt}\,.
  \label{qrel2}
\end{equation}
Then using Eqs.~\eqref{qrel} and~\eqref{qrel2}, we obtain
\begin{eqnarray}
  q(t) &=& - u \, {\rm Tr} \big[ \rho(0) \hat H_K(\mu) \big] \, e^{-ut}  
  \nonumber\\
  &=&
  - u \,\langle 0_{\mu_i} | \hat H_K(\mu) | 0_{\mu_i}\rangle \,e^{-ut} \,.
  \label{qrel3}
\end{eqnarray}
Note that 
\begin{equation}
  \langle 0_{\mu_i} | \hat H_K(\mu) | 0_{\mu_i}\rangle 
  = E_{0_{\mu_i}} +   W \,,
  \label{hkmui}
\end{equation}
where $E_{0_{\mu_i}} = \langle 0_{\mu_i} | \hat H_K(\mu_i) |
0_{\mu_i}\rangle$ is the ground-state energy at $\mu_i$, and $W$ is
the average work done at $t=0$, cf. Eq.~\eqref{woki}.  In particular
for $\bar\mu_i=0$
\begin{eqnarray}
\langle 0_{\mu_c} | \hat H_K(\mu) | 0_{\mu_c}\rangle &=&
L\, \left[4 D(L) - 1 + \bar\mu D(L) \right]
\label{hmu}\\
&=& L\left[ {\pi-4\over \pi}  - \bar\mu
 {\pi-2\over 2\pi}  + O(L^{-2})\right]
\,.
\nonumber
\end{eqnarray}
Equilibrium computations around the critical point give the general
structure~\cite{CPV-14}
\begin{equation}
  \langle 0_{\mu_i} | \hat H_K(\mu) | 0_{\mu_i}\rangle 
  = L g_a(\bar\mu_i,\bar{\mu}) + L^{-1} g_s(\kappa_i,\kappa) + O(L^{-2})\,.
  \label{eneav}
\end{equation}
Finally, we obtain
\begin{equation}
  Q(t) = \int_0^t dt \,q(t) = 
  \langle 0_{\mu_i} | \hat H_K(\mu) | 0_{\mu_i}\rangle 
  \left( e^{-ut} - 1 \right) \,.
  \label{qtf}
\end{equation}
The above results have been also carefully checked numerically, since
very accurate results for large lattice sizes [$L=O(10^3)$] can
be easily obtained exploiting the particular structure of the Lindblad
equations for the system considered~\cite{NRV-19, RV-19}.

Let us finally address the scaling behavior at the critical point
of the above results. In particular, we note that the average work
asymptotically behaves as
\begin{equation}
  W = L \,\bar\mu_i\,f_a(\bar\mu_i) + L^{-1} (\kappa_i-\kappa)
  f_s(\kappa_i) + O(L^{-2})\,.\quad
  \label{wsca0}
\end{equation}
Its structure does not apparently agree with the general scaling
behaviors put forward in Ref.~\cite{NRV-19-wo}. Indeed, taking into
account that $y_n=1$ is the renormalization-group dimension of the
perturbation involved by the quench of the parameter $\mu$, which is
the density operator $\hat n_x$, one would expect
\begin{equation}
  W \approx L^{-z} F_s(\kappa_i,\kappa)\,,
  \label{wsca}
\end{equation}
with $z=1$, which matches the subleading term in the
expansion~\eqref{wsca0}.  Equation~\eqref{wsca} is supposed to be the
asymptotic behavior keeping $\kappa_i$ and $\kappa$ fixed.  This
apparent contradiction is explained by contributions arising from
short-ranged fluctuations like those giving rise to the analytic part
of the free energy at the critical point~\cite{CPV-14}. In other
words, this is related to the mixing of the perturbation considered,
i.e. the particle density operator $\sum_x \hat n_x$, with the
identity operator, which leads to the leading term in
Eq.~\eqref{omun}. On the other hand, Ref.~\cite{NRV-19-wo} considered
perturbations not showing such problems, such as the longitudinal spin
operator in the quantum Ising chain. Generally, quenches associated
with perturbations vanishing at the critical point, by symmetry, give
rise to an average works satisfying the scaling laws reported in
Ref.~\cite{NRV-19-wo}.

Analogous considerations apply to the behavior of the heat interchange
around the critical point.  Indeed, using Eq.~\eqref{eneav}, we may
rewrite the expression of $q$, cf. Eq.~\eqref{qrel3}, in terms of
dynamic scaling variables~\eqref{dynscavarkii}, obtaining
\begin{equation}
  q(t) \approx L \,G_a(t,\mu_i,\mu,u) + 
  L^{-2} G_s(\theta,\kappa_i,\kappa,\gamma)\,,
  \label{qtsca}
\end{equation}
where the scaling part is again the subleading term
associated with the scaling function
\begin{equation}
  G_s(\theta,\kappa_i,\kappa,\gamma) = 
  -\gamma e^{-\gamma\theta}  g_s(\kappa_i,\kappa)\,.
  \label{scagpart}
\end{equation}
However, we should observe that disentangling the $O(L^{-2})$ scaling
part from the leading $O(L)$ term turns out to be a very hard task in
practice, due also to the fact that the different variables of the two
terms, with their different $L$ dependence, makes such distinction of
doubtful value.  Therefore, we do not pursue this issue further.

\end{document}